\documentclass[a4paper,fleqn]{cas-dc}

\usepackage[numbers, sort&compress]{natbib}
\usepackage{romannum}
\def\tsc#1{\csdef{#1}{\textsc{\lowercase{#1}}\xspace}}
\tsc{WGM}
\tsc{QE}
\begin{document}
\let\WriteBookmarks\relax
\def\floatpagepagefraction{1}
\def\textpagefraction{.001}

\pagenumbering{arabic}

\shorttitle{}    

\shortauthors{}  

\title [mode = title]{Exponential multi-graylevel computational-weighted dithering for high-quality binarized Fourier single-pixel imaging}  




\affiliation[1]{organization={CAS KeyLaboratory of Mechanical Behavior and Design of Materials, Department of Modern Mechanics, University of Science and Technology of China},
            addressline={}, 
            city={Hefei},
%          citysep={}, % Uncomment if no comma needed between city and postcode
            postcode={230027}, 
            state={Anhui},
            country={China}}

\affiliation[2]{organization={State Key Laboratory of Nonlinear Mechanics, Institute of Mechanics, Chinese Academy of Sciences},
            addressline={15 Beishihuan WestRoad}, 
            city={Beijing},
%          citysep={}, % Uncomment if no comma needed between city and postcode
            postcode={100190}, 
            state={Beijing},
            country={China}}

\affiliation[3]{organization={State Key Laboratory of Fire Science, University of Science and Technology of China},
            addressline={96 Jinzhai Road}, 
            city={Hefei},
%          citysep={}, % Uncomment if no comma needed between city and postcode
            postcode={230026}, 
            state={Anhui},
            country={China}}
            
\author[1]{Qigao Zhu}%[<options>]

% \cormark[1]


% \ead{}

% \ead[url]{https://orcid.org/0009-0000-1000-410X}

\credit{Conceptualization, Data curation, Formal analysis, Investigation, Validation, Visualization, Writing– original draft, Writing– review \& editing}

\author[1]{Haojia Jiang}%[]

% Footnote of the second author

% % Email id of the second author
% \ead{}


\credit{Writing– review \& editing}

\author[1]{Guan Wang}%[]
\cormark[1]
\credit{Writing– review \& editing, Investigation, Validation, Funding acquisition, Supervision}

\author[1]{Lianhao Zhang}%[]
\credit{Formal analysis}

\author[1]{Hanlei Gong}%[]
\credit{Formal analysis}

\author[1,2]{Huaxia Deng}%[]
\cormark[1]
\credit{Conceptualization; Funding acquisition, Supervision, Writing– review \& editing}

\author[1,3]{Xinglong Gong}
\cormark[1]
\credit{Funding acquisition, Project administration}


\cortext[1]{Corresponding author at: CAS Key Laboratory of Mechanical Behavior and Design of Materials, Department of Modern Mechanics, University of Science and Technology of China, Hefei, 230027, Anhui, China.}

\cortext[0]{Email addresses: \href{mailto:wangguan@ustc.edu.cn}{wangguan@ustc.edu.cn} (G.Wang), \href{mailto:hxdeng@ustc.edu.cn}{hxdeng@ustc.edu.cn} (H. Deng); \href{mailto:gongxl@ustc.edu.cn}{gongxl@ustc.edu.cn} (X. Gong) } 

% \fntext[0]{Email addresses: hxdeng@ustc.edu.cn (H. Deng), gongxl@ustc.edu.cn (X. Gong).}


\begin{abstract}
Binarized Fourier single-pixel imaging (FSI) takes full advantage of the high modulation speed of digital micromirror devices by applying Floyd-Steinberg spatial dithering to binarize grayscale Fourier patterns. However, the use of the spatial dithering introduces substantial quantization errors, leading to decreasing imaging quality. Here, we propose a binarization method for grayscale Fourier patterns based on exponential multi-graylevel computational‑weighted dithering, aimed at reducing quantization errors and then enhancing the imaging quality of binarized FSI. The proposed method quantizes Fourier patterns into $2^{R}$ values $\{0, 1/(R-1), 2/(R-1)... 1\}$ and then decomposes them into binarized patterns. Both simulation and experimental results demonstrate that the method significantly reduces quantization errors in Fourier coefficients acquisition and improves imaging quality. The mean absolute percentage error of Fourier coefficients decreases from 194\% to 28\% and the structural similarity of reconstructed images ($256\times256$ pixels) improves from 0.430 to 0.971, a 126\% enhancement compared to the conventional method. Lateral resolution of this proposed method almost approaches the theoretical lateral resolution limit calculated by Rayleigh Criterion.
\end{abstract}


% \begin{highlights}
% \item 
% \item 
% \item 
% \end{highlights}

%\nocite{*}

% Keywords

\maketitle

% Main text

\section{Introduction}
% Single-pixel imaging (SPI) acquires images using a single-pixel detector, operating on principles fundamentally distinct from conventional multi-pixel sensor array-based systems. \cite{B.Sun,WangGuan,Edgar,Peng,WenjingZhao} It offers advantages such as high detection sensitivity and widely applicable spectrum, making it commonly used in various fields \cite{YeJunTian-APL-2023}. 
Single-pixel imaging (SPI) is a computational optical imaging method whose operating principles are fundamentally distinct from conventional multi-pixel sensor array-based systems \cite{B.Sun,WangGuan,Edgar,Peng,WenjingZhao}. Instead of capturing a spatial image directly, SPI acquires scene information in the temporal dimension by sequentially modulating the target with a series of patterns and measuring the total reflected or transmitted light intensity with a single-pixel detector (typically a photodetector/PD rather than a CCD) \cite{JHeo_quantum-secured-single-pixe_APLPho, Edgar, LiuYujin_APL}. The PD offers high responsivity and a wide spectral range, making SPI a competitive solution for spectral bands where array detectors are unavailable or challenging, such as in terahertz imaging \cite{She, YangXingye_APL}.
This unique methodology grants SPI inherent advantages in detection sensitivity, spectral adaptability, and cost-effectiveness \cite{She}. The imaging process involves illuminating the scene with multiple modulation patterns and computationally reconstructing the image from the corresponding single-pixel measurements \cite{KatzOri_APL, ZhangZibang-NC-2015, ZhangZibang-OE-2017}. This approach has demonstrated broad application potential spanning fields such as microscopy \cite{Peng}, imaging through scattering media \cite{XinLU_ScattingMedia_AdvImaging}, quantum-secure communications \cite{JHeo_quantum-secured-single-pixe_APLPho}, and non-line-of-sight imaging \cite{GMusarra_Non-Line-of-Sight_PhysRevApplied}.

Fourier single-pixel imaging (FSI) \cite{ZhangZibang-NC-2015}, a key branch of this technology, uses Fourier basis patterns to acquire the Fourier spectrum of an object image. The image can be reconstructed by simply applying an inverse Fourier transform to the spectrum acquired. However, the Fourier basis patterns are inherently gray scale, which results in the difficulty that the patterns can hardly be generated at a high speed by using a commonly used high-speed spatial light modulator–digital micromirror device (DMD). Although it enables high-speed pattern generation and has high light utilization efficiency, a DMD can only generate one binarized pattern at a time due to its reflective nature. A DMD can synthesize a gray scale pattern through pulse width modulation, but it takes a much longer time than that of binarized pattern generation. Specifically, a typical DMD can generate over 20,000 binarized patterns per second, but can only generate 250 8-bit grayscale patterns per second. \cite{JiaxiangLi-OL-2023}

Many efforts have been made to improve the modulation speed of FSI. Zhang et al. proposed a binarized FSI approach that used spatial dithering to binarize the Fourier basis patterns \cite{ZhangZibang-SciR-2017}. Specifically, the Fourier patterns were first upsampled through interpolation to double their size and then binarized using the Floyd–Steinberg error-diffusion algorithm \cite{Floyd-Steinberg-Proc-1976, ZhangZibang-SciR-2017}. While this approach fully utilized the modulation speed of DMD, it reduced spatial resolution by merging 2 × 2 pixels into superpixels and introduced substantial quantization errors due to spatial dithering. Huang et al. introduced a signal dithering method that decomposed a Fourier pattern into a cluster of binarized patterns and assigned corresponding weights to the measurements \cite{Huang-OE-2018}. This method enhanced spatial resolution with a quantization level of 4 but quadrupled the sampling time. A lower quantization level (such as 2) may compromise this performance and introduce residual stripe artifacts in the image. The cost-effectiveness of this method still needs to be improved. Fan et al. proposed a positive-negative dithering method that applyed spatial dithering with a serpentine path directly to both positive and negative components of Fourier patterns before binarization \cite{Fan-OL-2025}. By quantizing these components into \{-1, 0, +1\} values and subsequently mapping them to binarized patterns, this method reduced quantization errors in Fourier coefficient acquisition. However, this method could still be further refined in handling high-frequency Fourier patterns, as some image details may not be fully retained due to residual quantization errors.

In this letter, we propose a Fourier pattern binarization method based on exponential multi-graylevel computational-weighted dithering to enhance the imaging quality of binarized FSI. Our approach is conceptually inspired by the success of multi-level halftoning and bit-plane decomposition techniques in digital image processing \cite{Digital-Halftoning-of-Images-1982, PIRSCH198331}, which have proven effective in representing continuous-tone images with limited output levels. We adapt and extend this concept to the specific problem of binarized FSI, where the goal is to minimize quantization errors in the Fourier domain.
Unlike the positive-negative dithering method \cite{Fan-OL-2025} that applies a single-level dithering, or the signal dithering method \cite{Huang-OE-2018} that decomposes a grayscale pattern into a cluster of equally-weighted binary patterns, our method introduces a multi-stage, computationally-weighted framework. The key novelty lies in: (1) the intermediate multi-graylevel representation, which captures more grayscale information before binarization, fundamentally reducing quantization errors compared to direct 2-level dithering; and (2) the exponentially-weighted decomposition of this intermediate pattern into binary patterns, where the weights (1/3, 2/3) are mathematically derived from a binary expansion, ensuring lossless reconstruction of the multi-graylevel pattern from the measurements. This combined strategy allows for a more faithful acquisition of the Fourier spectrum, especially its high-frequency components. The term ‘exponential’ in our method refers to the exponential number of intermediate graylevels $(2^R)$ and the binary (power-of-two) weighting scheme $\{2^0/2^R-1, 2^1/2^R-1, 2^2/2^R-1... 2^(R-1)/2^R-1\}$ used in the decomposition. This is analogous to the exponential weighting in binary number systems, where each bit position carries a weight that is a power of two. This design is fundamental to achieving a balanced trade-off between quantization accuracy and the number of projection patterns.
% Unlike the positive-negative dithering method \cite{Fan-OL-2025} that only applied dithering to grayscale Fourier patterns to generate positive and negative binarized patterns or the signal dithering method \cite{Huang-OE-2018} that directly decomposed a grayscale Fourier pattern into a cluster of binarized patterns and assigned corresponding weights to the measurements, our method utilizes the characteristics of the exponential weight by first quantizing these patterns into $2^{R}$ values $\{0, 1/(R-1), 2/(R-1)... 1\}$ and subsequently decomposing the multi-graylevel patterns to binarized patterns.These binarized patterns are used to illuminate the imaged object. The corresponding detected light intensities are then multiplied by their respective weighted decomposition coefficients and summed to reconstruct the Fourier spectrum. Finally, the object image is retrieved by applying an inverse Fourier transform.

% utilizes the characteristics of the Exponential weight by first performing multi-graylevel spatial dithering on the grayscale Fourier patterns and then decomposing the multi-graylevel patterns to binarized patterns.

\section{Method}

\begin{figure*}[ht]
\centering
\includegraphics[width=\linewidth]{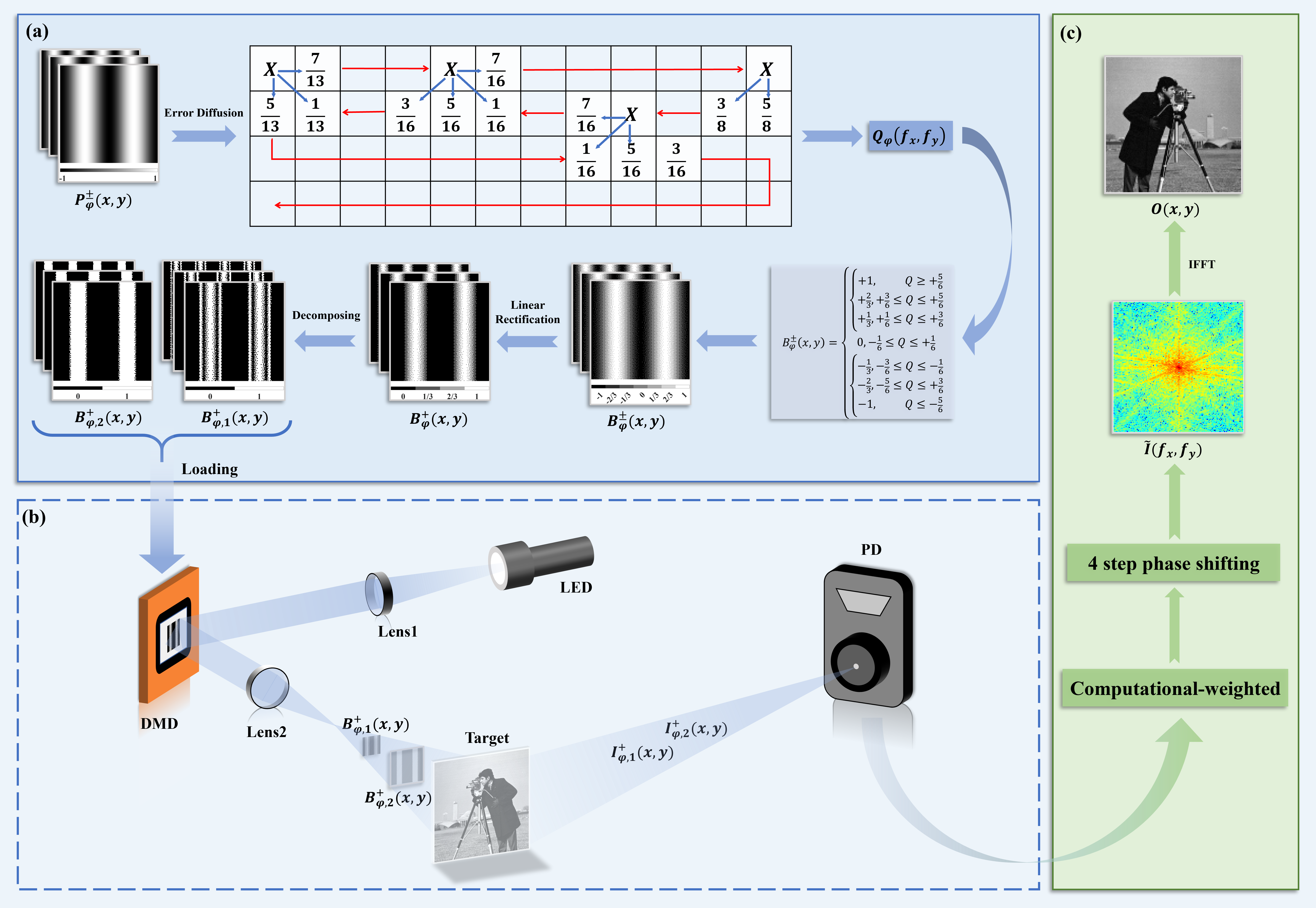}
% \begin{captionbox}
\caption{\label{fig1}Schematic diagram of exponential multi-graylevel dithering method for high-quality binarized Fourier single-pixel imaging. (a) Multi-graylevel dithering and binarized pattern generation. (b) Patterns projection and intensities acquisition. (c) Computational-weighted sum of detected intensities and reconstruction of object's image.}
% \label{fig1}
% \end{captionbox}
\end{figure*}

In FSI, an object’s spatial information is encoded by projecting Fourier basis patterns via a spatial light modulator (SLM). The resulting modulated light is subsequently collected by a single-pixel detector. A typical Fourier basis pattern is defined by an initial phase $\varphi$ and a spatial frequency pair $(f_{x}, f_{y})$:
\begin{equation}
\begin{aligned}
P_{\varphi}^{\pm}(x, y \mid f_{x}, f_{y}) &= \cos(2\pi(f_{x}x + f_{y}y) + \varphi).
% P_{\varphi}^{+}(x, y \mid f_{x}, f_{y}) &= \max(0, P_{\varphi}^{\pm}(x, y \mid f_{x}, f_{y})).
\end{aligned}
\label{1}
\end{equation}
% Considering that trigonometric induction formula, we can derive the following:
% \begin{equation}
% \begin{aligned}
% P_{0}^{\pm}(x, y \mid f_{x}, f_{y}) &= -P_{\pi}^{\pm}(x, y \mid f_{x}, f_{y}), \\
% P_{\pi/2}^{\pm}(x, y \mid f_{x}, f_{y}) &= -P_{3\pi/2}^{\pm}(x, y \mid f_{x}, f_{y}).
% \end{aligned}
% \label{2}
% \end{equation}
% The Fourier spectrum $\tilde{I}\left(f_{x}, f_{y}\right)$ is obtained using the four-step phase-shifting method as follows:\cite{Fan-OL-2025}
% \begin{equation}
% \begin{aligned}
% &\tilde{I}\left(f_{x}, f_{y}\right) \\ 
% & = \alpha\left(\sum_{x}\sum_{y}O(x,y)\left(P_{0}^{+}\left(x, y \mid f_{x},f_{y}\right)-P_{\pi}^{+}\left(x, y \mid f_{x},f_{y}\right)\right)\right.\\
% &\left.+j\sum_{x}\sum_{y}O(x,y)\left(P_{\pi/2}^{+}\left(x,y \mid f_{x},f_{y}\right)-P_{3\pi/2}^{+}\left(x,y \mid f_{x},f_{y}\right)\right)\right)
% \end{aligned}
% \label{3}
% \end{equation}
% where $\alpha$ is a constant factor related to the magnification of a single pixel detector and the spatial relationship between the detector and the object, and $O(x,y)$ represents the object's image. 
As shown in Fig.\ref{fig1}(a), we prounce $Q_{\varphi}(x,y \mid f_{x},f_{y})$ means the new gray patterns after error diffusion:
\begin{equation}
Q_{\varphi}(x,y \mid f_{x},f_{y}) = P_{\varphi}^{\pm}(x,y \mid f_{x},f_{y}) + \sum E(x,y),
\end{equation}
where $\sum E(x,y)$ represents the sum of errors diffused from other pixels. Besides, we apply a serpentine path \cite{Liang_2019} for our exponential multi-graylevel spatial dithering and modify the error diffusion kernel for pixels located at the beginning and end of rows. The method begins by applying multi-graylevel spatial dithering to transform the grayscale Fourier patterns $P_{\varphi}^{\pm}\left(x,y \mid f_{x},f_{y}\right)$ to the multi-graylevel patterns $B_{\varphi}^{\pm}(x,y \mid f_{x},f_{y})$, as defined in Eq.\ref{4}:
% \begin{equation}
% B_{\varphi}^{\pm}(x,y \mid f_{x},f_{y}) = 
% \begin{cases}
% +1, &Q \geq +\frac{5}{6}, \\
% +\frac{2}{3}, & +\frac{3}{6}\leq Q \leq +\frac{5}{6}\\
% +\frac{1}{3}, & +\frac{1}{6}\leq Q \leq +\frac{3}{6}, \\
% 0, & -\frac{1}{6}\leq Q \leq +\frac{1}{6}, \\
% -\frac{1}{3}, & -\frac{3}{6}\leq Q \leq -\frac{1}{6}, \\
% -\frac{2}{3}, & -\frac{5}{6}\leq Q \leq -\frac{3}{6}, \\
% -1, & Q \leq -\frac{5}{6}. \\
% \end{cases}
% \label{4}
% \end{equation}
\begin{equation}
B_{\varphi}^{\pm}(x,y\mid f_{x}, f_{y}) = 
\begin{cases}
+1, &Q \geq +\frac{2^{R+1}-3}{2^{R+1}-2}, \\
...\\
+\frac{i+1}{2^{R}-1}, & +\frac{2i+1}{2^{R+1}-2}\leq Q \leq +\frac{2i+3}{2^{R+1}-2}\\
+\frac{i}{2^{R}-1}, & +\frac{2i-1}{2^{R+1}-2}\leq Q \leq +\frac{2i+1}{2^{R+1}-2}\\
...\\
+\frac{1}{2^{R}-1}, & +\frac{1}{2^{R+1}-2}\leq Q \leq +\frac{3}{2^{R+1}-2}, \\
0, & -\frac{1}{2^{R+1}-2}\leq Q \leq +\frac{1}{2^{R+1}-2}, \\
-\frac{1}{2^{R}-1}, & -\frac{3}{2^{R+1}-2}\leq Q \leq -\frac{1}{2^{R+1}-2}, \\
...\\
-\frac{i}{2^{R}-1}, & -\frac{2i+1}{2^{R+1}-2}\leq Q \leq -\frac{2i-1}{2^{R+1}-2}, \\
-\frac{i+1}{2^{R}-1}, & -\frac{2i+3}{2^{R+1}-2}\leq Q \leq -\frac{2i+1}{2^{R+1}-2}, \\
...\\
-1, & Q \leq -\frac{2^{R+1}-3}{2^{R+1}-2}, \\
\end{cases}
\label{4}
\end{equation}
where $R$ is the number of intermediate graylevels, and $i$ is from 0 to $2^{R}-1$. The resulting multi-graylevel patterns $B_{\varphi}^{\pm}(x,y\mid f_{x}, f_{y})$ have $2^{R+1}-1$ graylevels. We experimentally investigated the impact of using different $R$ (e.g., 1, 2, 3). Our simulations in Fig.S1 of \href{Supplement Material.pdf}{supplementary material} showed that $R=2$  provides the optimal balance between reconstruction quality and time cost. Using fewer levels led to visible contouring artifacts, while using more levels offered diminishing returns in quality at the cost of increased projection time. When $R=2$, the multi-graylevel pattern $B_{\varphi}^{\pm}(x,y \mid f_{x},f_{y})$ could be defined as:
\begin{equation}
B_{\varphi}^{\pm}(x,y \mid f_{x},f_{y}) = 
\begin{cases}
+1, &Q \geq +\frac{5}{6}, \\
+\frac{2}{3}, & +\frac{3}{6}\leq Q \leq +\frac{5}{6}\\
+\frac{1}{3}, & +\frac{1}{6}\leq Q \leq +\frac{3}{6}, \\
0, & -\frac{1}{6}\leq Q \leq +\frac{1}{6}, \\
-\frac{1}{3}, & -\frac{3}{6}\leq Q \leq -\frac{1}{6}, \\
-\frac{2}{3}, & -\frac{5}{6}\leq Q \leq -\frac{3}{6}, \\
-1, & Q \leq -\frac{5}{6}, \\
\end{cases}
\end{equation}
which is then processed by the linear rectification function as Eq.\ref{5}:
% (refer to chapter \newcounter{temp}\setcounter{temp}{1}\Roman{temp} of \href{Supplement Material.pdf}{supplementary material} for an analysis of the factors determining the specific number 7 of graylevel in the multi-graylevel patterns $B_{\varphi}^{\pm}(x,y \mid f_{x},f_{y})$, as well as the rationale behind the specific weight 1/3 and 2/3 used in subsequent contexts),
\begin{equation}
B_{\varphi}^{+}(x, y \mid f_{x},f_{y}) = \max\left(0, B_{\varphi}^{\pm}(x, y \mid f_{x},f_{y})\right).
\label{5}
\end{equation}
As shown in Fig.\ref{fig1}(a), the pattern $B_{\varphi}^{+}(x, y \mid f_{x},f_{y})$ is 4-graylevel patterns, as defined in Eq.\ref{6}: 
\begin{equation}
B_{\varphi}^{+}(x, y \mid f_{x},f_{y}) = \frac{1}{3}B_{\varphi,1}^{+}(x, y \mid f_{x},f_{y})+\frac{2}{3}B_{\varphi,2}^{+}(x, y \mid f_{x},f_{y}),
\label{6}
\end{equation}
where $B_{\varphi,1}^{+}(x, y \mid f_{x},f_{y})$ and $B_{\varphi,2}^{+}(x, y \mid f_{x},f_{y})$ are defined in Eq.\ref{8}:
\begin{equation}
\begin{aligned}
B_{\varphi,1}^{+}(x, y \mid f_{x},f_{y}) = \left\lfloor \frac{3B_{\varphi}^{+}(x, y \mid f_{x},f_{y})}{2^0} \right\rfloor \bmod 2 , \\
B_{\varphi,2}^{+}(x, y \mid f_{x},f_{y}) = \left\lfloor \frac{3B_{\varphi}^{+}(x, y\mid f_{x},f_{y})}{2^1} \right\rfloor \bmod 2 .
\end{aligned}
\label{8}
\end{equation}
The $\lfloor\cdot\rfloor $ is floor rounding operation and mod is modulo operation.

As shown in Fig.\ref{fig1}(b), then the binarized patterns $B_{\varphi,1}^{+}(x, y \mid f_{x},f_{y})$ and $B_{\varphi,2}^{+}(x, y \mid f_{x},f_{y})$ could be projected on the target to acquire the convolution $I_{\varphi}^{+}(f_{x}, f_{y})$ of pattern and imaging object:
\begin{equation}
\begin{aligned}
&I_{\varphi}^{+}\left(f_{x}, f_{y}\right) \\
&= \sum_{x, y} \alpha  B_{\varphi}^{+}\left(x, y \mid f_{x},f_{y}\right) O(x,y) \\
&= \sum_{x, y}\left[\frac{\alpha}{3} B_{\varphi, 1}^{+}\left(x, y \mid f_{x}, f_{y}\right) +\frac{2\alpha}{3} B_{\varphi, 2}^{+}\left(x, y \mid f_{x}, f_{y}\right)\right]O(x,y) \\
&= \frac{1}{3}\sum_{x, y}\alpha B_{\varphi, 1}^{+}\left(x, y \mid f_{x}, f_{y}\right) O(x,y)\\
&+\frac{2}{3}\sum_{x, y}\alpha B_{\varphi, 2}^{+}\left(x, y \mid f_{x}, f_{y}\right) O(x,y) \\
&= \frac{1}{3} I_{\varphi, 1}^{+}\left(f_{x}, f_{y}\right)+\frac{2}{3} I_{\varphi, 2}^{+}\left(f_{x}, f_{y}\right),
\end{aligned}
\label{9}
\end{equation}
where the parameters $I_{\varphi, 1}^{+}\left(f_{x}, f_{y}\right)$ and $I_{\varphi, 2}^{+}\left(f_{x}, f_{y}\right)$ are the detected intensities corresponding to the binarized illumination patterns $B_{\varphi, 1}^{+}\left(x, y \mid f_{x}, f_{y}\right)$ and $B_{\varphi, 2}^{+}\left(x, y \mid f_{x}, f_{y}\right)$ interacting with the imaged object, respectively. The constant factor $\alpha$ is related to the illumination time, the magnification of a single pixel detector and the spatial relationship between the detector and the object. Eq.\ref{9} shows that the weighted sum of the detected intensities for the binarized illumination patterns interacting with the imaged object can be equivalent to the intensities for the 4-graylevels patterns $B_{\varphi}^{+}\left(x, y \mid f_{x}, f_{y}\right)$ interacting with the imaged object.

As shown in Fig.\ref{fig1}(c), for each $\left(f_{x}, f_{y}\right)$, the convolution value $I^{+}_{\varphi}(f_{x},f_{y})$ could be acquired by calculating the weighted sum of detected intensities as shown in Eq.\ref{9}. The 4-step shifting is used to acquire the Fourier spectogram $\tilde{I}\left(f_{x}, f_{y}\right)$ and then the object’s image $O(x,y)$ is reconstructed via the IFFT (Inverse Fast Fourier Transform) as shown in Eq.\ref{10}: 
\begin{equation}
\begin{aligned}
    O(x,y)
    &=\mathcal{F}^{-1}\{\tilde{I}\left(f_{x}, f_{y}\right)\}  \\
    &=\mathcal{F}^{-1}\{I_{0}^{+}(f_{x}, f_{y})-I_{\pi}^{+}(f_{x}, f_{y})\\
    &+j(I_{\pi/2}^{+}(f_{x}, f_{y})-I_{3\pi/2}^{+}(f_{x}, f_{y}))\}.
\end{aligned}
\label{10}
\end{equation}

% The details of math theory can be found in the chapter 1 of the \href{Supplement Material.pdf}{supplementary material}.

\section{Simulation}

% \subsection{Simulation}

A number of simulations were conducted to validate our method. The simulation was performed by using a city-gate image cropped from the DIV2K database \cite{Timofte-CVPRW-2017} and resized to $256\times256$ pixels as the target image.

\begin{figure*}[htbp]
\centering
\includegraphics[width=0.9\linewidth]{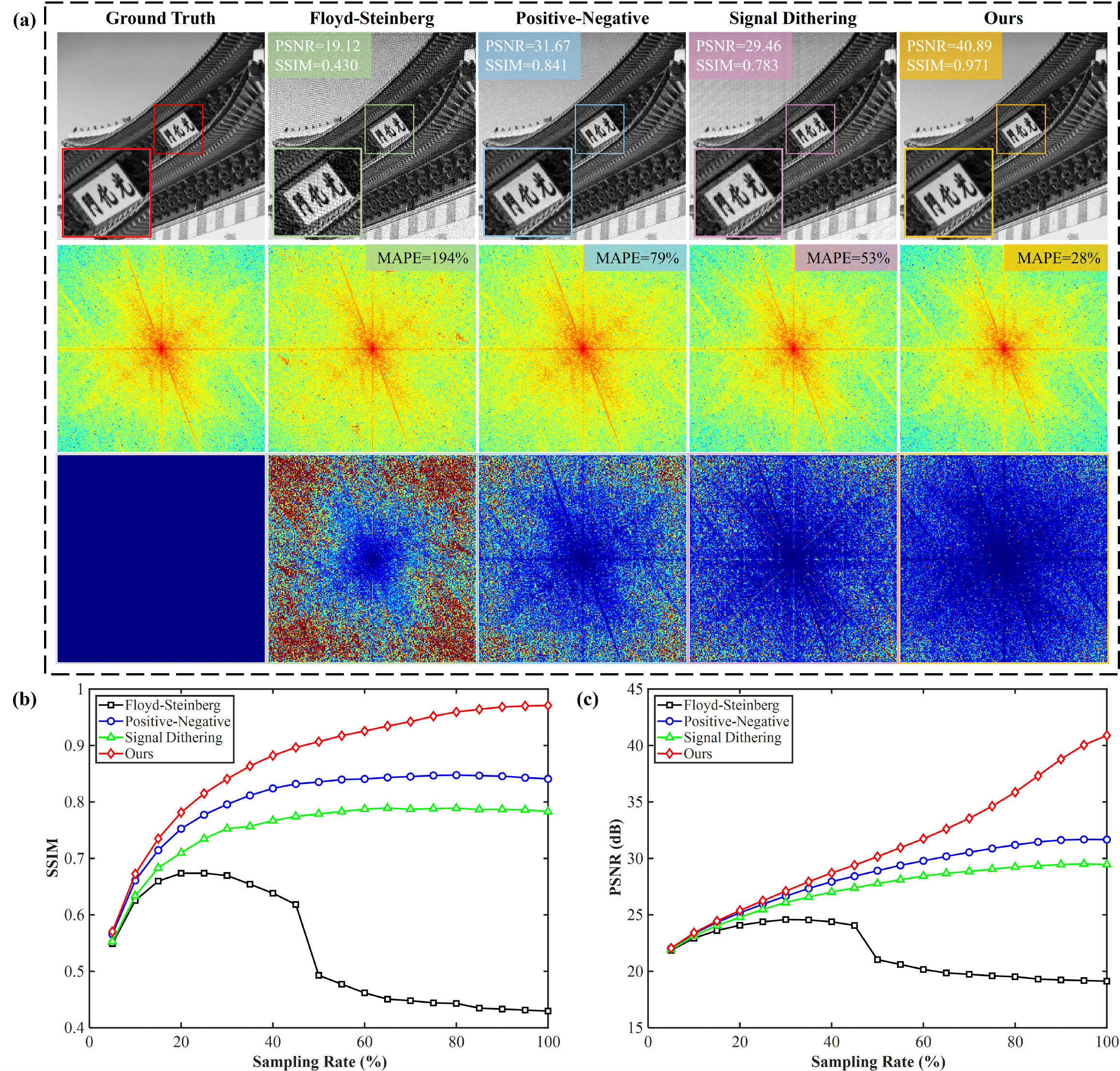}
% \subfloat[]{\includegraphics[width=0.75\linewidth]{figure2.png}\label{fig2a}}\\
% \subfloat[]{\includegraphics[width=0.38\linewidth]{figure2-b.png}\label{fig2b}} \hspace{5pt}
% \subfloat[]{\includegraphics[width=0.38\linewidth]{figure2-c.png}\label{fig2c}}\\
% \subfloat[]{\includegraphics[width=0.76\linewidth]{figure16.png}\label{fig2d}} 
% \begin{captionbox}
\caption{City-gate image ($256\times256$ pixels) and simulation results via four methods (including Floyd-Steinberg algorithm\cite{Floyd-Steinberg-Proc-1976, ZhangZibang-SciR-2017}, signal dithering algorithm\cite{Huang-OE-2018}, positive-negative dithering algorithm\cite{Fan-OL-2025}, and our method) at 100\% sampling rate, along with the corresponding PSNRs and SSIMs. (a) Reconstructed images (first row), spectrograms (second row) and the absolute percentage error of spectrograms (third row) of 4 dithering methods. (b)-(c) SSIM and PSNR curves of four dithering methods under different sampling rates.}
% \end{captionbox}
\label{fig2}
\end{figure*}

We compared the reconstruction results of our method with other dithering algorithms,including the Floyd–Steinberg algorithm \cite{Floyd-Steinberg-Proc-1976, ZhangZibang-SciR-2017}, the signal dithering algorithm \cite{Huang-OE-2018} with a quantization level of 2, and the positive-negative dithering algorithm \cite{Fan-OL-2025}. Two full-reference image quality metrics were used: structural similarity (SSIM) \cite{ZhouWang-TIP-2004} and peak signal-to-noise ratio (PSNR).
As shown in Fig.\ref{fig2}(a), our method achieved the highest reconstruction quality, with less noise and clearer local details (highlighted in the close-up of Fig.\ref{fig2}(a)). Our PSNR and SSIM values significantly outperformed those of the other methods. On average, our method achieved a PSNR value more than 10 dB greater, and an SSIM value over 0.1 greater, than the other methods. 

Here, the sampling rate is defined as the ratio of the number of actually acquired Fourier coefficients to the total number of coefficients required for a full Nyquist-rate reconstruction (256×256 = 65,536 coefficients). Note that due to the two-pattern decomposition, our method projects twice as many binary patterns as the conventional Floyd-Steinberg method for each acquired coefficient. We analyzed the PSNR and SSIM values of the four methods across different sampling rates. The Fourier patterns were sampled in a circular order \cite{ZhangZibang-OE-2017}, and 20 sampling rates were evaluated in this simulation. As shown in Fig.\ref{fig2}(b), the SSIM of the images from the other three methods decreased as the sampling rate increased, indicating that greater quantization errors were introduced in the Fourier patterns at higher spatial frequencies. In contrast, our method maintained the highest reconstruction quality across all sampling rates. As shown in Fig.\ref{fig2}(c), we can see that our method achieves robust performance across the entire range from low to high sampling rates. Especially, our method could acquire SSIM over 0.97 and PSNR over 40 dB at 100\% sampling rate. An interesting observation is the performance degradation of the Floyd-Steinberg dithering method at high sampling rates ( > 50\%). This counter-intuitive behavior can be explained by the nature of the error-diffusion algorithm. At high spatial frequencies, the accumulated quantization errors from the dithering process become spatially correlated and manifest as structured noise artifacts in the reconstructed Fourier spectrum. When more high-frequency coefficients are sampled, these artifacts accumulate, leading to a net decrease in reconstruction quality. In contrast, our proposed method effectively suppresses these errors, allowing the image quality to continuously improve with more samples.

To further investigate, we also compared Fourier spectrograms of 4 methods as shown in Fig.\ref{fig2}(a). The simulation results show the mean absolute percentage error (MAPE) index value of Fourier coefficients acquired by our method is only 28\%, compared with 194\% of Floyd-Steinberg dithering, 79\% of positive-negative dithering and 53\% of signal dithering. As shown in Fig.\ref{fig2}(a), our method is able to reduce quantization errors in Fourier coefficients acquisition especially in the high-frequency region. To ensure the generality of our findings, we extended our simulations to include several other test images (tower, girl, peacock) from the DIV2K dataset. The average PSNR and SSIM values across these images are presented in Table S1 in Supplementary Material, consistently demonstrating the superior performance of our method.
% Additional tests on other images, presented in the chapter \setcounter{temp}{2}\Roman{temp} of \href{Supplement Material.pdf}{supplementary material}, further validated our method’s superior performance.
% To further investigate, we simulated reconstructed images at low sampling rates (see the chapter 2 of \href{aipguide4-1.pdf}{supplementary material}). The results show that at super subNyquist sampling rates ($<$10\%), the performance gap of the four methods becomes narrow, but as the sampling rate increases, the performance gap widens, with our method consistently yielding better results. 
% We also examined the reconstructed image quality of the four methods with the same pattern count (see Fig. 3). The simulation results indicate that our method could achieve the highest image quality with 2*M*N sampling times when M equals 256 and N equals 256. These results substantiate that our method achieves higher imaging quality to the other dithering methods under ultra-low sampling regimes while significantly reducing acquisition time through a binarized modulation scheme that better aligns with the DMD operational mechanism.

\section{Experiment}

\begin{figure*}[ht]
\centering
\includegraphics[width=\linewidth]{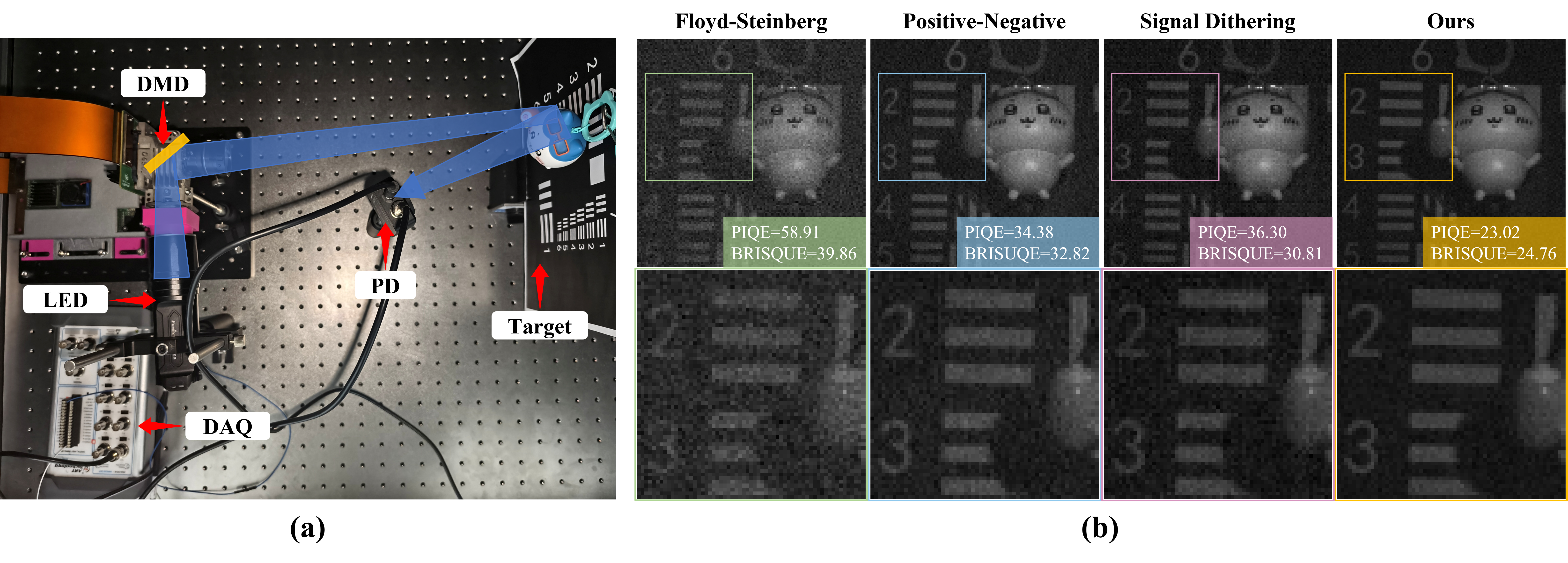}
\caption{Imaging of a toy: (a) Experimental setup with top-down view. DMD: digital micromirror device; LED: light emitting diode; DAQ: data acquisition board; PD: photodetector. (b) Comparison results of our method with the other three dithering methods. Image size: $128\times128$ pixels. Sampling Rate: 100\%. The bottom row shows magnified views of selected regions. }
\label{fig4}
\end{figure*}

\begin{figure*}
    \centering
    \includegraphics[width=0.9\linewidth]{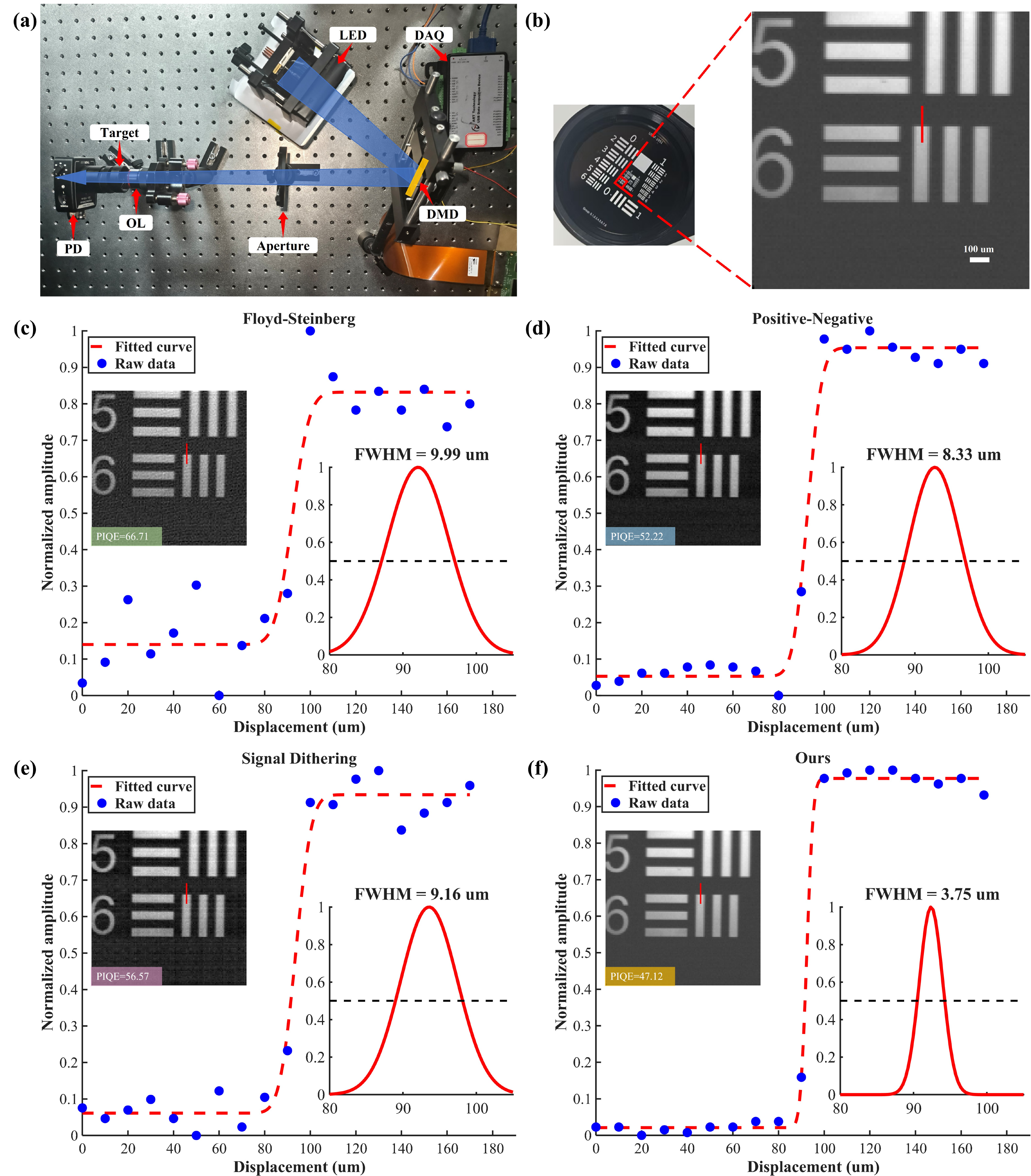}
    \caption{(a) Experimental setup with top-down view. DMD: digital micromirror device; LED: light emitting diode; DAQ: data acquisition board; PD: photodetector; OL: $4\times$ objective lens. (b) Image of the sharp edge in a USAF resolution target.(Scale bars, 100 µm.) (c-f) Normalized edge spread function measured at the focal plane using a sharp edge on the USAF target. The corresponding line spread functions were fitted to compute the lateral resolutions defined by the full-width at half-maximum (Inset).}
    \label{fig6}
\end{figure*}

Experimental validation further confirms the effectiveness of the proposed dithering strategy in enhancing the imaging quality of FSI. As shown in Fig.\ref{fig4}(a), the setup comprises a DMD (DLPV-7001) with a white light source, DAQ (PCIe-9770), and a photodetector (PD, Thorlabs PDA100A2). Patterns were projected at 4 kHz. See the chapter \Romannum{4} of \href{Supplement Material.pdf}{supplementary material} for other imaging targets.

Experimental results are shown in Fig.\ref{fig4}(b). Our method was compared with the other three dithering algorithms, demonstrating consistency between experimental and simulation results. The image reconstruction was performed by applying the IFFT directly to the computational-weighted detected intensities, with no further image post-processing applied. The test target was a toy. Our method achieved significant noise reduction in the reconstructed images. The enlarged regions of the image clearly demonstrate the superiority of our proposed method. The other three methods exhibit noticeable artifacts, while our method could reconstruct significantly a more realistic scene. In order to conduct quantitative evaluation, we need to find new quality assessment metrics that are suitable for the reconstructed images. Given the challenges in acquiring reference images with consistent viewpoints and magnification factors, we employed the no-reference image quality assessment metrics BRISQUE \cite{BRISQUE-TIP-2012} and PIQE \cite{PIQE-NCC-2015}. Both BRISQUE and PIQE exhibit strong correlations with human subjective evaluations, where lower numerical values correspond to higher visual quality. 

For the test target (toy), our method performs the best compared to the other three methods in terms of both PIQE score and BRISQUE under full sampling, validating the method’s efficacy and broad applicability in enhancing imaging quality. Specifically, our method can reduce the PIQE and BRISQUE metrics to 23.02 and 24.76 respectively, while the two metrics of other methods are over 30. To further investigate, additional experiment results at low sampling rates are provided in the chapter \Romannum{3} of \href{Supplement Material.pdf}{supplementary material}. The supplementary content shows that at extremely low sampling rates, the performance gap of the four methods is narrow, but as the sampling rate increases, the performance gap widens, with our method yielding better results. 

The above experimental results indicate that the proposed method can obtain images with more details and less noise than the other three methods, and has great potential for microscopic imaging with a near-diffraction-limit optical resolution while the other three methods may not approach the diffraction limit. This could be verified through microscopic imaging experiment. As shown in Fig.\ref{fig6}(a), the experimental setup for microscopic imaging comprises a DMD (DLPV-7001) with a white light source, a DAQ (PCIe-9770), an objective lens (4×, NA 0.1, 160/0.17), an aperture and a photodetector (PD, Thorlabs PDA100A2). Patterns were projected at 4 kHz. 

The edge spread function (ESF) describes how an ideal optical system images a perfectly sharp edge. The line spread function (LSF) is the first derivative of ESF. The full-width at half-maximum (FWHM) of LSF is a direct, quantitative measure of the optical system's lateral resolution. In essence, a smaller FWHM indicates higher lateral resolution. Besides, PIQE is selected as the evaluation metric for USAF microscopy imaging results because it can well describe the noise level of the image.

This setup in Fig.\ref{fig6}(a) is to measure the lateral resolution of 4 methods and Fig.\ref{fig6}(b) shows the image of a sharp edge on the USAF target, from which the ESF was measured. Its corresponding LSF is fitted to compute the lateral resolution which is defined by the full-width at half-maximum of the LSF. As shown in Fig.\ref{fig6}(c)-(f), the lateral resolution of 4 methods is measured. As shown in Fig.\ref{fig6}(f), our method's imaging result has the least PIQE value (47.12) while this metric of other methods is over 50, indicating that our method has the least amount of noise. The lateral resolution of our method is 3.75 um which almost approaches the theoretical lateral resolution limit of 3.4 um calculated by Rayleigh Criterion $0.61\lambda/N_{A}$. Overall, these results establish that our method is capable of microscopic imaging with a near-diffraction-limit optical resolution due to reducing quantization errors in Fourier coefficients acquisition especially in the high-frequency region.

\section{Application}

\begin{figure*}[ht]
\centering
\includegraphics[width=\linewidth]{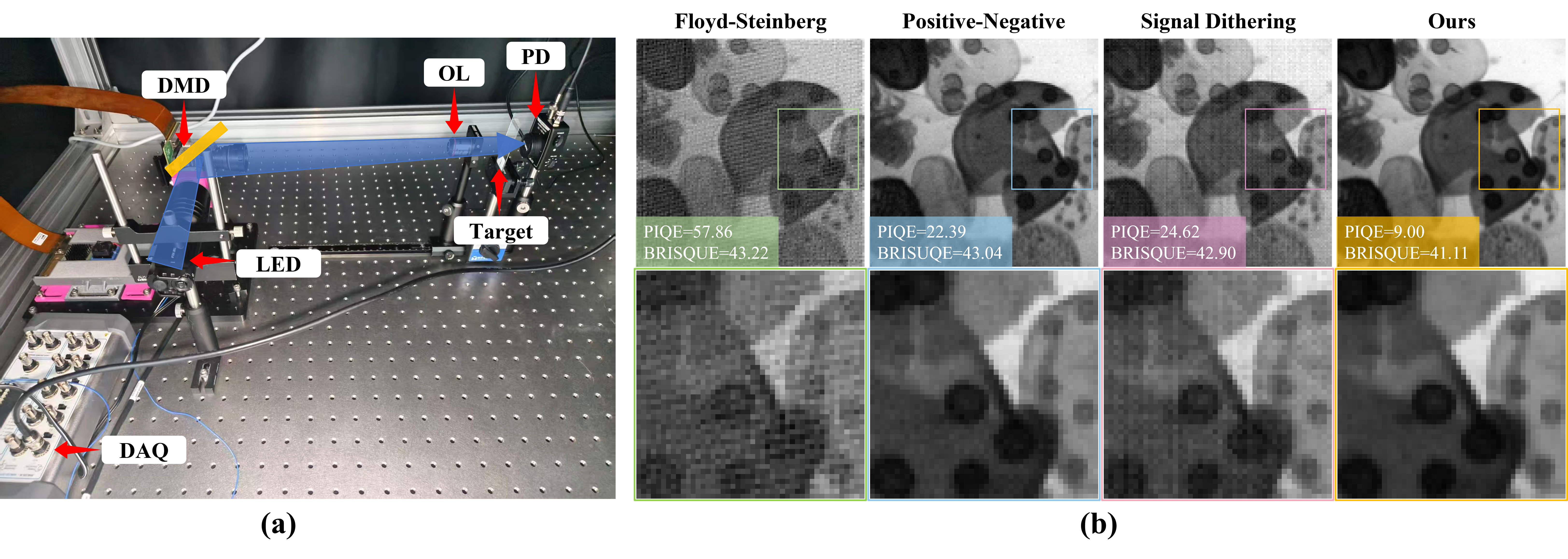}
% \subfloat[]{\includegraphics[width=0.395\linewidth]{figure15-b.pdf}\label{fig7a}} \hfill
% \subfloat[]{\includegraphics[width=0.595\linewidth]{figure6.png}\label{fig7b}}
\caption{Imaging of volvox: (a) Experimental setup with top-down view. DMD: digital micromirror device; LED: light emitting diode; DAQ: data acquisition board; PD: photodetector; OL: $4\times$ objective lens. (b) Comparison results of our method with the other three dithering methods. Image size: $128\times128$ pixels. Sampling Rate: 100\%. The bottom row shows magnified views of selected regions. }
\label{fig7}
\end{figure*} 

The above experimental results indicate that our method has provided positive advancements for the application of microscopic imaging. Therefore, the proposed method will be applied to microscopic imaging of Volvox cells and compared with three other methods. 

The setup shown in Fig.\ref{fig7}(a) is used for microscopic imaging of volvox cells, and it includes a DMD (DLPV-7001) with a white light source, DAQ (PCIe-9770), an objective lens (4×, NA 0.1, 160/0.17) and a photodetector (PD, Thorlabs PDA100A2). Patterns were projected at 4 kHz.

As shown in Fig.\ref{fig7}(b), the microscopic imaging result of Floyd-Steinberg dithering exhibits severe noise, while that of signal ditherig displays pronounced vertical and horizontal stripes. Positive-negative dithering, despite being free from evident noise and stripes, still suffers from blurring in certain details and edges. However, our proposed method further addresses the issues present in positive-negative dithering, enabling the acquisition of remarkably clear details and sharp edges. Overall, the microscopic imaging results indicate that the proposed method is more suitable for the application scenario of microscopic imaging compared to the other three methods.

\section{Discussion and Conclusion}

In summary, we proposed a grayscale pattern binarization method for binarized FSI, significantly improving image quality, thereby preserving the fine image details. Unlike previous approaches that directly applied spatial dithering to grayscale Fourier patterns to generate 2-graylevel patterns, namely binarized patterns, our method utilizes the characteristics of the exponential weight by first performing multi-graylevel spatial dithering on the grayscale Fourier patterns and then decomposing the multi-graylevel patterns to binarized patterns. 

We demonstrated the effectiveness of this method through both simulations and experiments. The MAPE of Fourier coefficients decreases to 28\%, showing that our method significantly reduces quantization errors introduced by dithering. The superiority of our method is evident compared to existing techniques such as positive-negative dithering \cite{Fan-OL-2025} and signal dithering \cite{Huang-OE-2018}. In simulated reconstructions (256 × 256 pixels), our method yields SSIM of 0.971 and PSNR of 40.89 dB, surpassing the performance of positive-negative dithering (0.841, 31.67 dB) and signal dithering (0.783, 24.96 dB). In the case of undersampling, the imaging quality of our method can continuously improve with the increase of sampling rate, while the imaging quality of other methods remains almost unchanged or even decreases at high sampling rates (>50\%). This further indicates that our method can reduce the quantization errors of Fourier coefficients in the high-frequency region. Experimental results further confirm the practical advantages of our method. For the imaging result of toy, our method yields the best scores (PIQE: 23.02, BRISQUE: 24.76), outperforming Floyd-Steinberg dithering (58.91, 39.86), positive-negative dithering (34.38, 32.82), and signal dithering (36.30, 30.81). This proposed method achieves a lateral resolution of 3.75 um which almost approaches the theoretical lateral resolution limit of $4\times$ objective lens (3.4 um).

It is important to note that the proposed method achieves this significant improvement in image quality at the cost of doubling the number of projection patterns required per Fourier coefficient (two patterns instead of one). Consequently, for a fixed DMD refresh rate (4 kHz in our experiments), the total acquisition time is doubled compared to the standard Floyd-Steinberg dithering method. However, this trade-off is highly favorable for applications prioritizing image quality over acquisition speed, such as high-resolution microscopic imaging. Furthermore, the DMD’s inherent high-speed binary modulation capability (over 20,000 patterns/sec) ensures that even with patterns doubling, the total acquisition time remains manageable for many practical scenarios. Substantially, this method enables high-quality image reconstruction in binarized FSI experiments without requiring additional image post-processing, providing positive advancements for applications such as fast single-pixel imaging, single-pixel video imaging\cite{DENG2025112552}, motion tracking single-pixel imaging \cite{Zhang:24}, and single-pixel microscopic imaging.

\printcredits

\section*{Declaration of Competing Interest}
The authors declare that they have no known competing financial interests or personal relationships that could have appeared to influence the work reported in this paper.

\section*{Acknowledgments}
This work was financially supported by the National Natural Science Foundation of China (Grant Nos. 12427802, 12372187, 12132016, 52321003, and 12502222), CAS Talent Introduction Program (No. KJ2090007006), Fundamental Research Funds for the Central Universities (No. WK2480000010 and WK2090250108), China Postdoctoral Science Foundation (No. 2024M763151), Postdoctoral Fellowship Program of CPSF (No. GZB20250466). 

\section*{Appendix A. Supplementary material}
See the \href{Supplement Material.pdf}{supplementary material} for comprehensive technical details, including the theoretical basis of the exponential multi-graylevel computational-weighted dithering algorithm, additional simulation results for other images, additional experimental results in under-sampled cases, and additional imaging targets.

\section*{Data availability}
The data that support the findings of this study are available from the corresponding author upon reasonable request.

% \bibliographystyle{cas-model2-names}
\bibliographystyle{unsrtnat}

\bibliography{references}



\end{document}